\documentclass[]{spie}  

\usepackage{amsmath,amsfonts,amssymb}
\usepackage{graphicx}
\usepackage[colorlinks=true, allcolors=blue]{hyperref}

\title{Feedhorn development and scalability for Simons Observatory and beyond}

\author[a]{Sara M. Simon}
\author[a]{Joseph E. Golec}
\author[b]{Aamir Ali}
\author[c]{Jason Austermann}
\author[c]{James A. Beall}
\author[d]{Sarah Marie M. Bruno}
\author[d]{Steve K. Choi}
\author[d]{Kevin T. Crowley}
\author[e]{Simon Dicker}
\author[c]{Bradley Dober}
\author[c]{Shannon M. Duff}
\author[d]{Erin Healy}
\author[b,f]{Charles A. Hill}
\author[d]{Shuay-Pwu Patty Ho}
\author[c]{Johannes Hubmayr}
\author[d]{Yaqiong Li}
\author[e]{Marius Lungu}
\author[a]{Jeff McMahon}
\author[e]{John Orlowski-Scherer}
\author[g]{Maria Salatino}
\author[d]{Suzanne Staggs}
\author[h]{Edward J. Wollack}
\author[e]{Zhilei Xu}
\author[e]{Ningfeng Zhu}

\affil[a]{Department of Physics, University of Michigan, Ann Arbor, MI, USA}
\affil[b]{Department of Physics, University California, Berkeley, Berkeley, CA, USA}
\affil[c]{Quantum Sensors Group, NIST, Boulder, CO, USA}
\affil[d]{Department of Physics, Princeton University, Princeton, NJ, USA}
\affil[e]{Department of Physics \& Astronomy, University of Pennsylvania, Philadelphia, PA, USA}
\affil[f]{Physics Division, Lawrence Berkeley National Laboratory, Berkeley, CA, USA}
\affil[g]{AstroParticle and Cosmology (APC) laboratory, Paris Diderot University, Paris, France}
\affil[h]{NASA/Goddard Space Flight Center, Greenbelt, MD, USA}

\authorinfo{Send correspondence to S.M.S.\\S.M.S.: E-mail: smsimon@umich.edu}

\begin{document}
\maketitle

\begin{abstract}
The Simons Observatory (SO) will measure the cosmic microwave background (CMB) in both temperature and polarization over a wide range of angular scales and frequencies from 27-270~GHz with unprecedented sensitivity. One technology for coupling light onto the $\sim$50 detector wafers that SO will field is spline-profiled feedhorns, which offer tunability between coupling efficiency and control of beam polarization leakage effects. We will present efforts to scale up feedhorn production for SO and their viability for future CMB experiments, including direct-machining metal feedhorn arrays and laser machining stacked Si arrays.
\end{abstract}


\section{Introduction}
\label{sec:intro}  
The cosmic microwave background (CMB) temperature and polarization anisotropies are a precise tool for probing several fundamental questions, including the number and masses of neutrinos, the nature of dark energy and dark matter, and if there was a period of inflation in the early universe~\cite{s4_science}. However, the CMB polarization is faint and can be contaminated by foregrounds from synchrotron and dust emission, so future CMB experiments will require high sensitivity, wide frequency coverage, and minimized systematic effects. The Simons Observatory (SO) will observe the CMB temperature and polarization with unprecedented sensitivity and provide critical technological advances for the next-generation of CMB experiments, CMB-S4~\cite{s4_science,s4_tech}. A 6~m crossed-Dragone telescope will enable measurements at small angular scales to constrain the properties of neutrinos, dark energy, and dark matter while several $\sim$0.5~m refractive telescopes will observe large angular scales to probe inflation. SO will employ superconducting multichroic detectors with wide frequency coverage across six bands with band centers spanning from 27~GHz to 270~GHz. The detectors used in CMB experiments are photon noise limited, so increased sensitivity must come from increased detector numbers. While current CMB experiments have between $\sim$5,000 to $\sim$10,000 detectors, SO will have $>60,000$ detectors and CMB-S4 will require on order $\sim$500,000 detectors. This means producing and fielding more detector arrays than ever before. To scale technologies for use on SO and next-generation projects, the cost and production time of their manufacture must be significantly reduced.


A critical aspect of the detector arrays is
the technology used to couple light onto the pixels. One technology for this optical coupling is feedhorns. Feedhorns have been used broadly in sub-mm, CMB, and terahertz applications~\cite{Leech_2011,Leech_2012}. SO plans to field multichroic feedhorn-coupled detector arrays at 90/150~GHz and 220/270~GHz. Three types of feedhorns have been used in CMB experiments: conical feedhorns, corrugated feedhorns, and spline-profiled feedhorns. Conical feedhorns have high beam coupling efficiency, but have poor systematic performance. Corrugated feedhorns approach near-ideal beam symmetry, resulting low levels of beam systematic effects. However, in tightly-packed detector arrays, their corrugations represent a significant fraction of the available aperture, resulting in lower beam coupling efficiency~\cite{Simon_SPIE_2016}. Further, to produce multichroic corrugated feedhorns, ring-loaded structures are required in the corrugations. These ring-loaded features are difficult to fabricate, making them time consuming and costly to produce. For these reasons, we limit our studies to spline-profiled feedhorns. Spline-profiled feedhorns offer tunability between beam coupling efficiency and controlling beam systematic effects and can be made monotonically increasing in diameter, making them easier to fabricate in both Si and metal~\cite{Simon_SPIE_2016,Yassin_2007,Zeng_2010,Granet_2004}.

Feedhorn arrays for CMB applications are typically fabricated out of photolithographically-patterned, stacked Si wafers or direct-machined into Al. Using proven production methods and current tooling, photolithography limits the stacked Si array size to 150~mm, can be costly, and takes weeks to produce a single array. Direct-machined Al feedhorn arrays must contend with differential thermal contraction with the Si detector wafer and have less precise tolerances in their fabrication, which becomes particularly important at higher frequencies. We will present the feedhorn research and development efforts for SO and their viability for future CMB experiments. Section~\ref{sec:direct_machine} will discuss direct-machined feedhorn arrays, including developments in Al, Si-Al alloy, and Mo feedhorn arrays. Section~\ref{sec:silicon} will discuss Si array fabrication methods including photolithographic patterning and laser machining.

\section{Direct-Machined Feedhorn Arrays} \label{sec:direct_machine}
Metal feedhorn arrays are fabricated by direct-machining feedhorn profiles into the metal either with a custom drill and reamer set or with a bull nose cutter. This process allows for faster fabrication than stacked Si feedhorn arrays, and as such, is often lower cost. Because the full array is metal, it is robust and mounting features for readout and the detector array assembly can be directly machined into the array. The fabrication of direct-machined metal feedhorn arrays can be outsourced to a machine shop. This allows for increased production rates, which is a necessary step for next-generation CMB experiments. Direct-machined metal feedhorn arrays are slated for use on several current-generation CMB experiments, including Advanced ACTPol (AdvACT)~\cite{Simon_2018} and CLASS~\cite{CLASS_2014}.

However, there are several drawbacks to metal feedhorn arrays. Because all of the features are machined, higher tolerances in both the alignment and the feedhorn profiles themselves are harder to achieve, which becomes increasingly important at high frequencies. Differential thermal contraction between the metal feedhorn array and the Si detector array may also cause greater misalignment between these parts and can add higher mechanical risk from increased stresses. An ideal material for a direct-machined metal feedhorn array would have low density to minimize the array mass and a thermal contraction similar to that of Si. Metal feedhorn arrays, especially when they are made from superconducting materials, require the addition of a normal metal (e.g., Au, Cu, or other plating layer with low magnetic susceptibility) for heat sinking. Depending on the material, many standard plating chemistries require a base layer of Ni for adhesion. Because Ni is ferromagnetic, it can suppress the critical temperature of the transition-edge sensors and cause trapped flux in the readout, which is comprised of superconducting quantum interference devices. While the Ni base layer can be a concern, several experiments including AdvACT and CLASS have demonstrated that the effects of the Ni can be negligible. Ni can be avoided by using non-standard plating chemistries, but this limits the available vendors and can increase the cost. We discuss three different materials for direct-machined feedhorn arrays: Al 6061, Si-Al alloy, and Mo.

\subsection{Aluminum 6061}
Al~6061 is an inexpensive material and is easy to machine, allowing for a low rate of tool wear and fast production. A 90/150~GHz feedhorn array with 507~feedhorns for SO only takes $\sim$20~hours to produce and is currently less than 20\% of the cost of a Si platelet feedhorn array using photolithographic patterning. Because of this, SO will employ Al feedhorns for its 90/150~GHz feedhorn arrays. The Au coating we consider for the SO feedhorns uses a thin (0.5~$\mathrm{\mu}$m) Ni base layer for increased adhesion and thermal conductivity, but we note that non-ferromagnetic materials like Pd can be used with Al. For this study, we used a custom drill and finishing reamer set from Custom Cutting Tools\footnote{http://www.customcuttingtools.com/}, but a bull nose cutter could also be used.

\subsubsection{Feedhorn Spacing}
\begin{figure}[h!]
\centering
\includegraphics[width=0.9\textwidth]{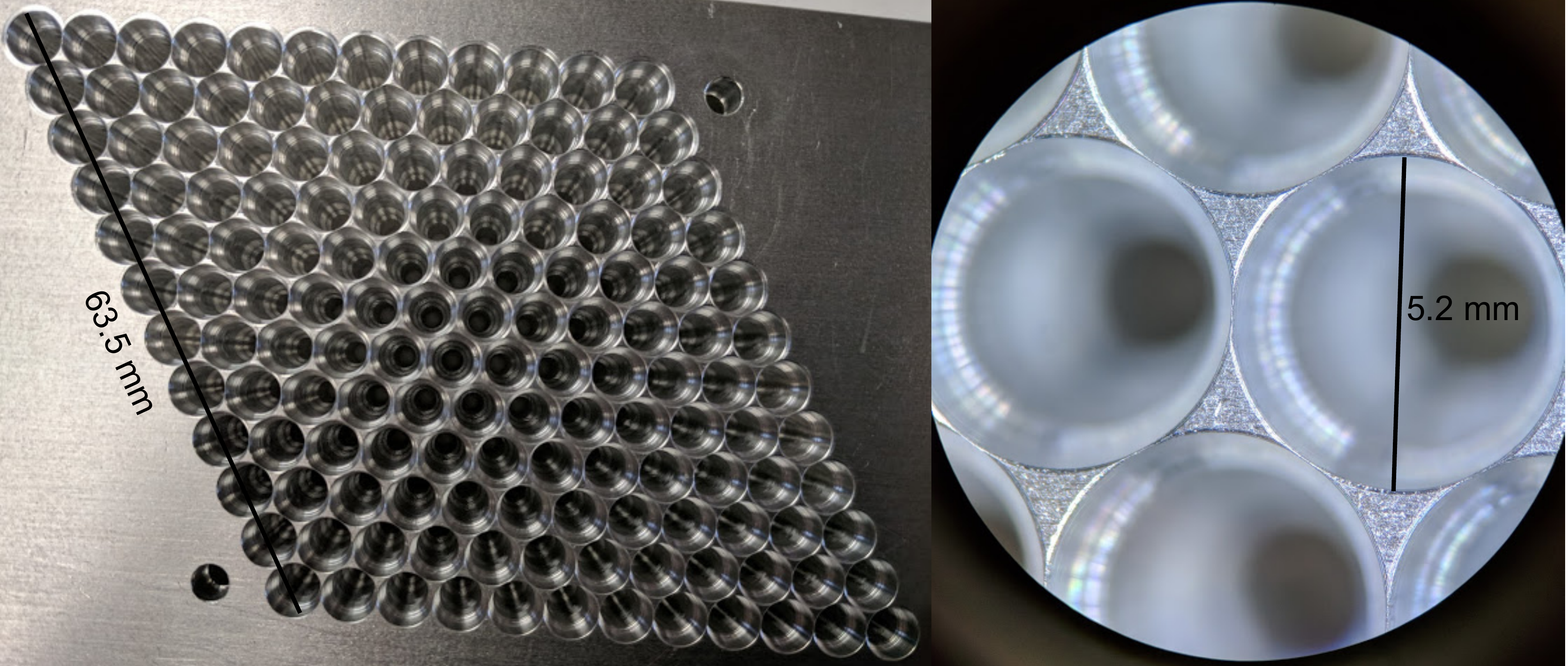}
\caption{A prototype Al~6061 90/150~GHz feedhorn array with 144~horns and a 5.3~mm horn pitch is shown above. The full array is shown on the left, and a zoomed in image is shown on the right. The horn sidewalls are 100~$\mathrm{\mu}$m, and there are no major defects. This represents 1/3 of a full SO feedhorn array. While 100~$\mathrm{\mu}$m horn sidewalls with a 5.3~mm feedhorn pitch are possible, they are not robust to pressure directly on the sidewall, so 150~$\mathrm{\mu}$m wide sidewalls are baselined for the SO 90/150~GHz feedhorn arrays.}
\label{fig:mini_array}
\end{figure}

\begin{figure}[h!]
\centering
\includegraphics[width=0.8\textwidth]{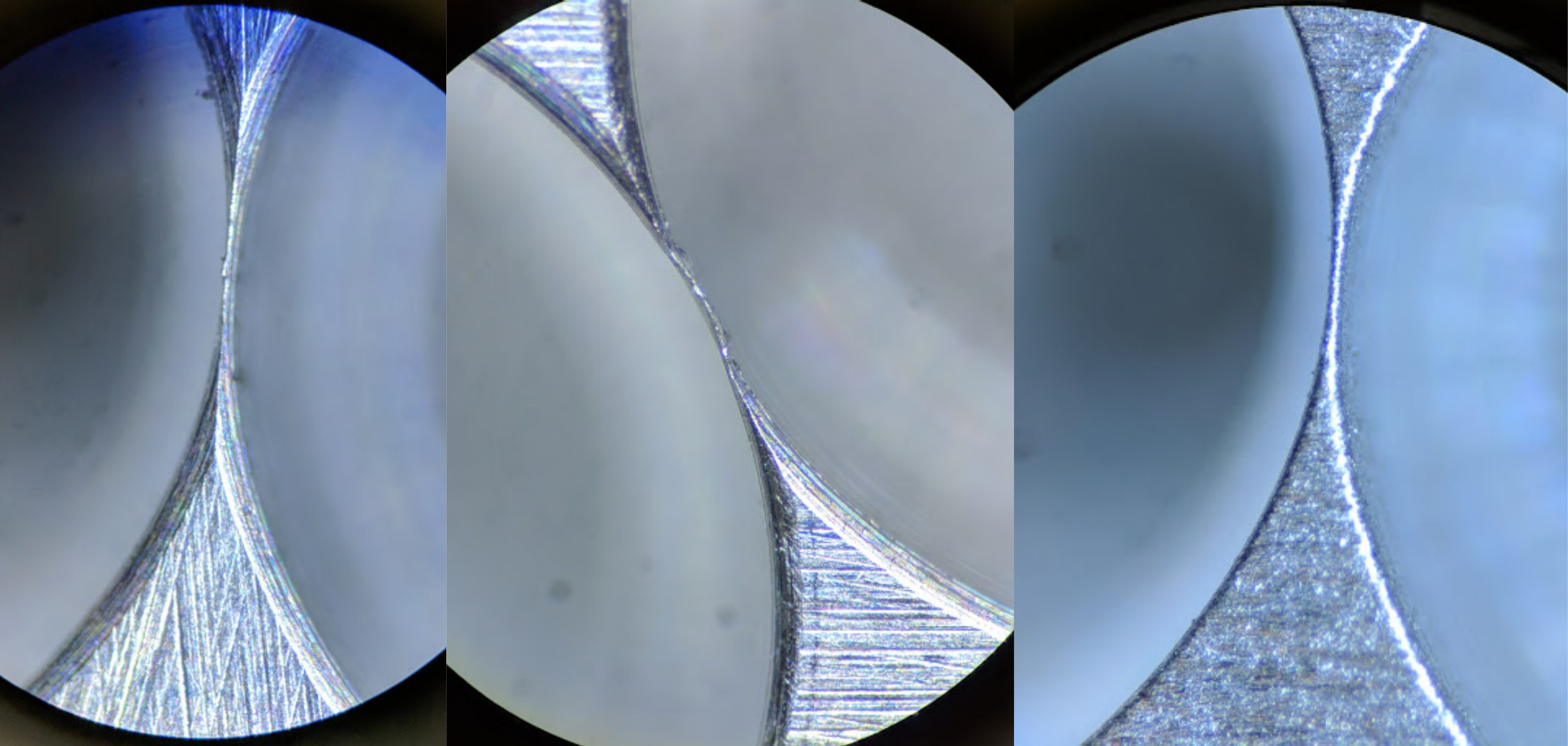}
\caption{Feedhorn sidewalls with a thickness of 100~$\mathrm{\mu}$m (left and center) and 150~$\mathrm{\mu}$m (right) are shown above. While a 100~$\mathrm{\mu}$m sidewall is possible for a 5.3~mm horn pitch, it is not robust to direct pressure on the sidewall, which can cause minor defects as shown in the center panel. A sidewall 150~$\mathrm{\mu}$m thick is robust against direct pressure and is baselined for the SO 90/150~GHz feedhorn array. As the distance between feedhorns decreases, so does the minimum robust wall thickness.}
\label{fig:feed_wall}
\end{figure}

One concern with metal feeds is the spacing between feedhorns. To maximize the feedhorn performance for a given pixel pitch, the aperture of the feedhorn must be maximized and thus the wall between feedhorns minimized. Si feedhorn arrays can achieve a sidewall thickness of 50~$\mathrm{\mu}$m but use a 100~$\mathrm{\mu}$m sidewall for robustness. For SO, we have demonstrated that a feedhorn array with 100~$\mathrm{\mu}$m spacing and 144 horns at a 5.3~mm pixel pitch can be machined with Al~6061 without any major defects (Figure~\ref{fig:mini_array}). The left panel of Figure~\ref{fig:feed_wall} 
shows a magnified view of the 100~$\mathrm{\mu}$m feedhorn wall. The 100~$\mathrm{\mu}$m sidewall array is robust to even pressure across the top of the array, but if direct pressure is applied to individual walls, they can bend or break. Additionally, 28\% of the walls in the array exhibit minor defects like that pictured in the center panel of Fig.~\ref{fig:feed_wall} though we expect these to have a negligible effect on the beam. We note that a harder Al may be more robust at the 100~$\mathrm{\mu}$m spacing but that there would be additional tool wear. The wall between feedhorns at a 5.3~mm pitch becomes robust against direct pressure at a thickness of 150~$\mathrm{\mu}$m (right panel of Fig.~\ref{fig:feed_wall}), which also eliminates the minor defects observed at the 100~$\mathrm{\mu}$m spacing. For a 90/150~GHz SO feedhorn with a pixel pitch of 5.3~mm, this decrease in aperture size from 5.2~mm to 5.15~mm results in a $\sim$1\% decrease in efficiency. Because the decrease in performance is small, the SO 90/150~GHz Al feedhorns will use the more sturdy 150~$\mathrm{\mu}$m spacing between horns. At smaller pixel pitches, a 100~$\mathrm{\mu}$m spacing becomes more robust, so thinner walls may be achievable for smaller pixel spacings.

\subsubsection{Tolerances}
Direct-machining cannot currently match the tolerances achievable with photolitography and laser machining. The accuracy of the feedhorn profiles, their locations in the array, and the alignment with the detector array are critical and become more important as frequency increases. We measure the tolerances of the feedhorns through non-contact metrology and measuring the beams of the feedhorns with a beam-mapper setup. We also model the misalignment between the feedhorns and the detector array in electromagnetic simulation software to set alignment requirements.

We use a non-contact metrology system to measure the profile of the feedhorns, the uniformity of the Au coating, the wall thicknesses, the distance between feedhorns, and any ellipticity in the apertures of the horns. The conoscopic sensor in the metrology system works by emitting a laser beam that reflects off of the object surface and back to the sensor. The beam is then passed through an optically anisotropic crystal and interfered with a reference beam to produce an interference pattern that is used to determine the distance from the surface.

\begin{figure}[h!]
\centering
\includegraphics[width=0.6\textwidth]{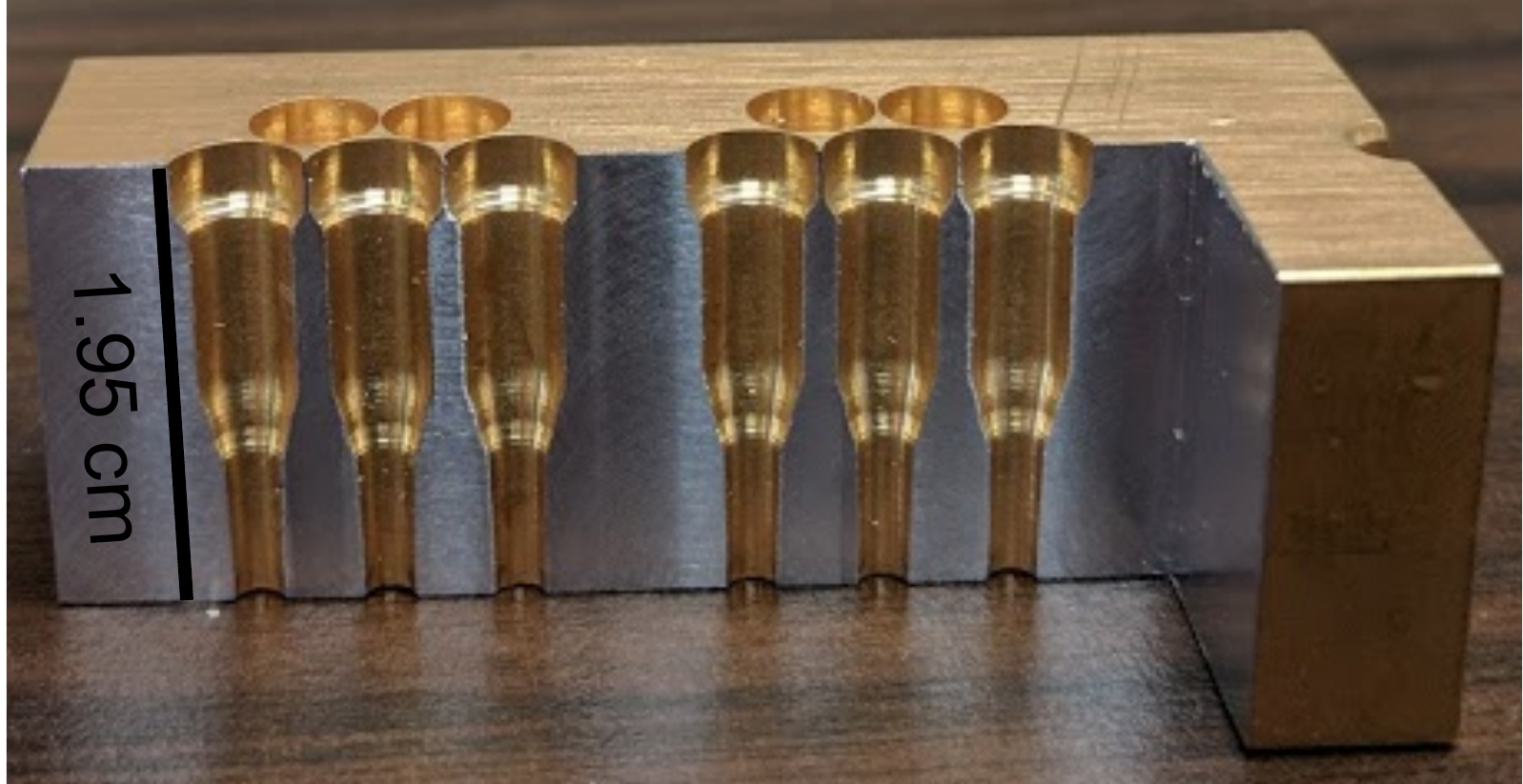}
\caption{A cross-section of the 90/150~GHz prototype array after Au coating is shown above. The cross-sections are measured down the center with a non-contact metrology probe to assess the accuracy of the feedhorn reamer set, the machining depth, and the Au coating uniformity. The measurements for this cross-section are shown in Fig.~\ref{fig:MF_gold}.}
\label{fig:MF_cross}
\end{figure}

\begin{figure}[h!]
\centering
\includegraphics[width=0.55\textwidth]{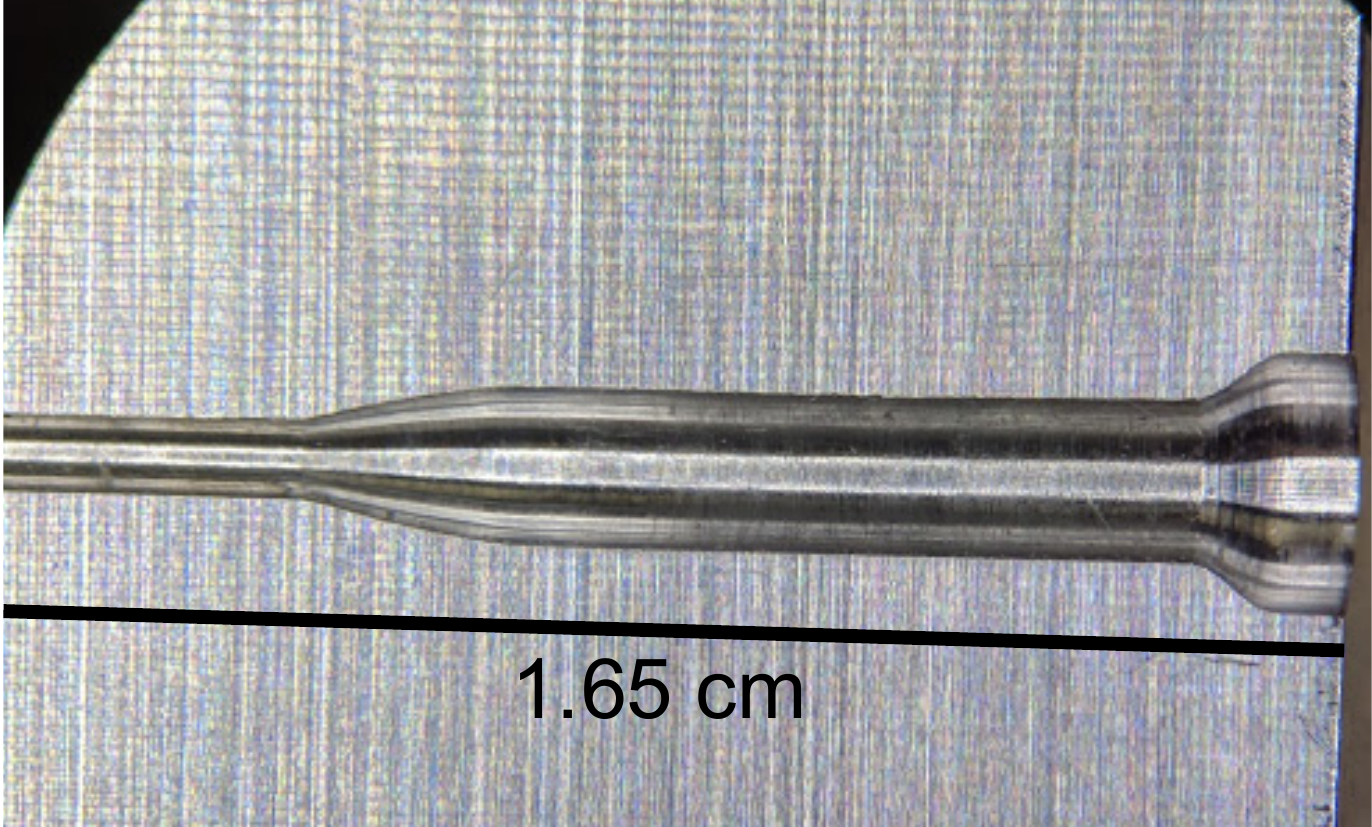}
\caption{A cross-section of a prototype 220/270~GHz feedhorn is shown above. The measurements of this cross-section are shown in Fig.~\ref{fig:UHF_profile}. Higher frequencies are more sensitive to tolerances in the profile and alignment.}
\label{fig:UHF_cross}
\end{figure}

\begin{figure}[h!]
\centering
\includegraphics[width=0.95\textwidth]{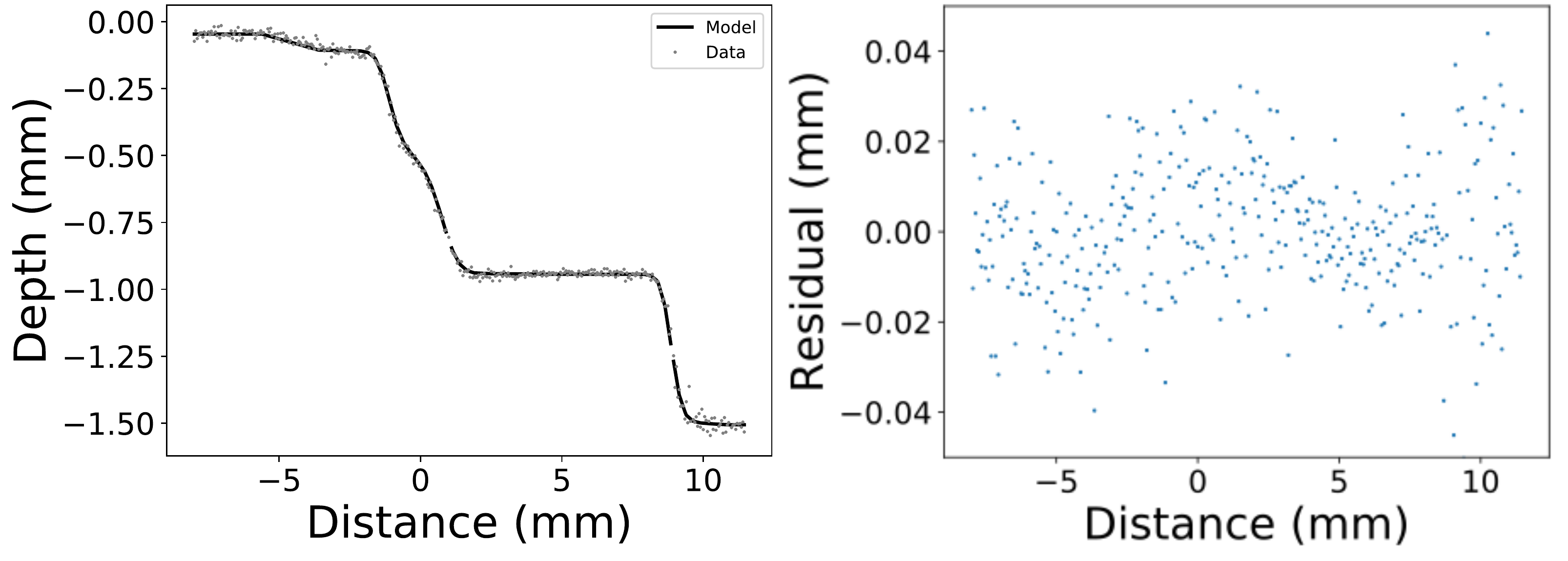}
\caption{Measurements of a cross-sectioned prototype 90/150~GHz feedhorn is shown above. The left figure shows the feedhorn design (black line) with its measured profile (grey points). The residuals are shown in the right plot. The RMS of the profile measurement is 17~$\mathrm{\mu}$m.}
\label{fig:MF_profile}
\end{figure}

\begin{figure}[h!]
\centering
\includegraphics[width=0.95\textwidth]{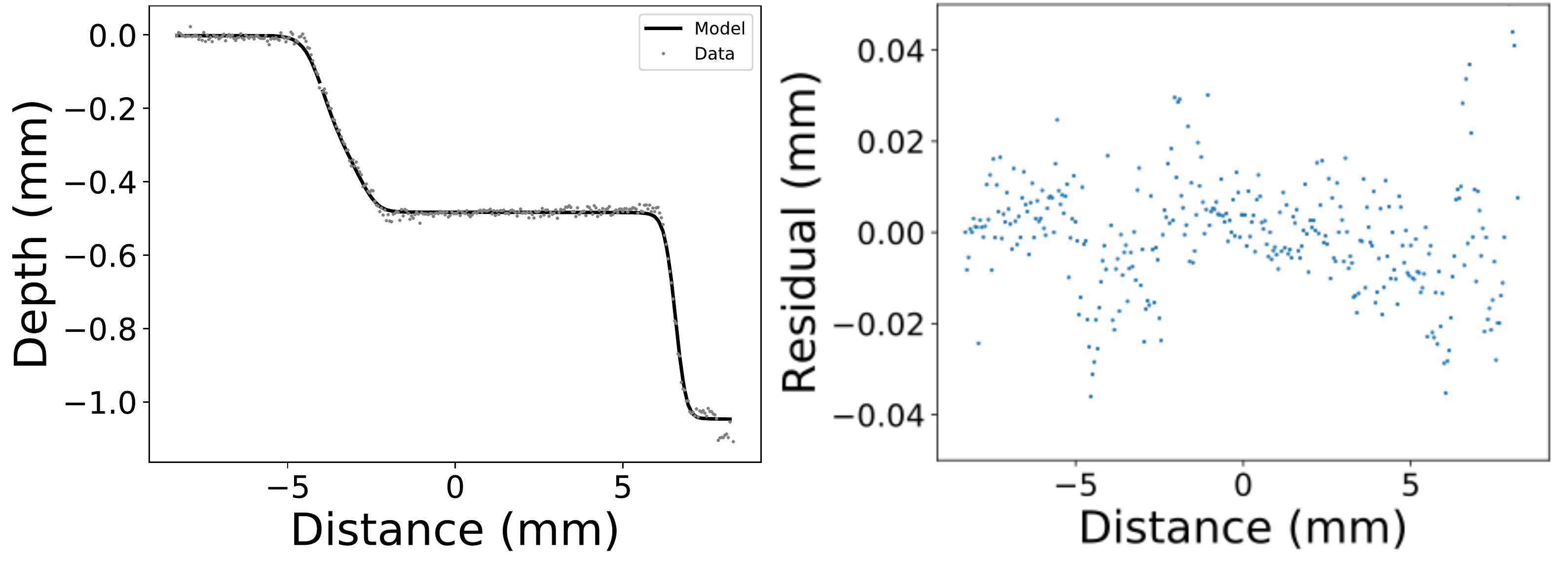}
\caption{Measurements and residuals of a cross-sectioned prototype 220/270~GHz feedhorn are shown above with the same convention as Fig~\ref{fig:MF_profile}. The cross-section measurements of the feed show that the reamer had a defect that can be seen on the far right portion of the plots. Measurements like this can be used to check that the reamer and machining tolerances meet the requirements prior to array production.}
\label{fig:UHF_profile}
\end{figure}

\begin{figure}[h!]
\centering
\includegraphics[width=0.95\textwidth]{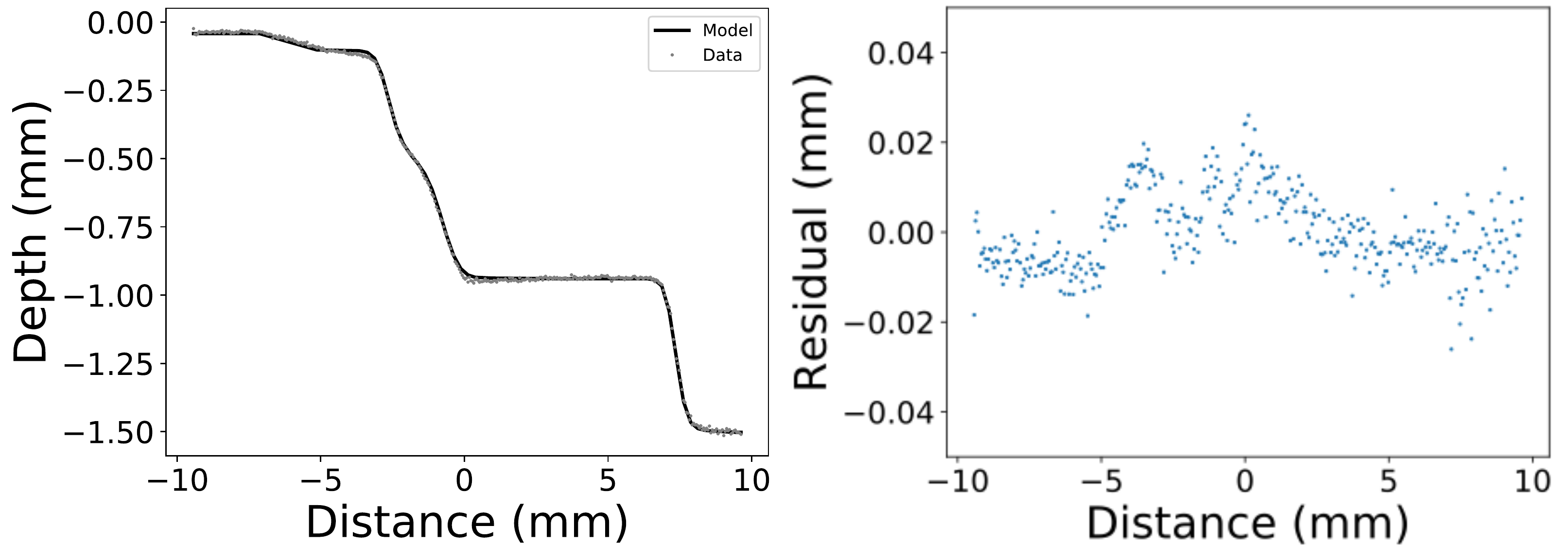}
\caption{Measurements and residuals of a cross-sectioned prototype 90/150~GHz feedhorn after Au coating are shown above with the same convention as Fig~\ref{fig:MF_profile}. It can be seen that the Au coating is uniform across the profile.}
\label{fig:MF_gold}
\end{figure}

Profile measurements can be used to verify that the reamer profile matches the design profile and to set the correct machining depth for the feedhorn profile before production. We measure the profiles of prototype 90/150~GHz and 220/270~GHz feedhorns by cross-sectioning a single feedhorn as shown in Fig.~\ref{fig:MF_cross} and~\ref{fig:UHF_cross}. The profile is measured down the center of the cross-section to avoid burrs from the machining during cross-sectioning. Figures~\ref{fig:MF_profile} and~\ref{fig:UHF_profile} show the measured feedhorn profiles with the designed feedhorn profile and their residuals. The root mean square (RMS) of the profiles is 17~$\mathrm{\mu}$m and 11~$\mathrm{\mu}$m for the 90/150~GHz and 220/270~GHz profiles, respectively. The uncertainty of the probe is $\sim$6~$\mathrm{\mu}$m but can increase if the material is highly reflective. Using this method, we were able to determine that the reamer for the 220/270~GHz profile had a production error, which can be seen in Fig.~\ref{fig:UHF_profile}. We also use profile cross-sections to assess the uniformity of a Au coating inside of the horns for heat sinking as shown in Fig.~\ref{fig:MF_gold} and find that it has an RMS of 8~$\mathrm{\mu}$m. This is a lower RMS than the uncoated Al feedhorn profile, which is likely due to the reduced reflectivity of the Au-coated profiles compared to bare Al. We will also input the RMS of the measured profiles into the feedhorn models to determine the scientific impact of the profile tolerances. Non-contact metrology will be used to vet the reamer set, machining tolerances, and Au coating for the final arrays before moving into full production.

The non-contact metrology setup can also be used to verify the feedhorn wall thickness and to ensure that the feedhorn profiles are not elliptical. 
Beam maps of the feedhorns are also underway to verify the feedhorn performance we expect based on the non-contact metrology measurements.


With $\Delta L/L=41.4\times 10^{-4}$, Al has the highest level of differential thermal contraction compared to Si ($\Delta L/L=2.2\times 10^{-4}$) of the materials discussed in this paper. To account for this, the feedhorns and their spacing are over-sized to account for the thermal contraction from room temperature to 100~mK. Simulations to set alignment requirements between the feedhorn and detector arrays are underway along with tests to determine if the arrays meet the requirements.

These tolerancing tests will determine if Al~6061 is a viable option for the SO 220/270~GHz arrays for SO and next-generation CMB experiments. Demonstration of these feedhorns at 90/150~GHz and possibly 220/270~GHz on SO will be a critical step forward for future Al feedhorn production.

\subsection{Silicon-Aluminum Alloy}
\begin{figure}[h!]
\centering
\includegraphics[width=0.7\textwidth]{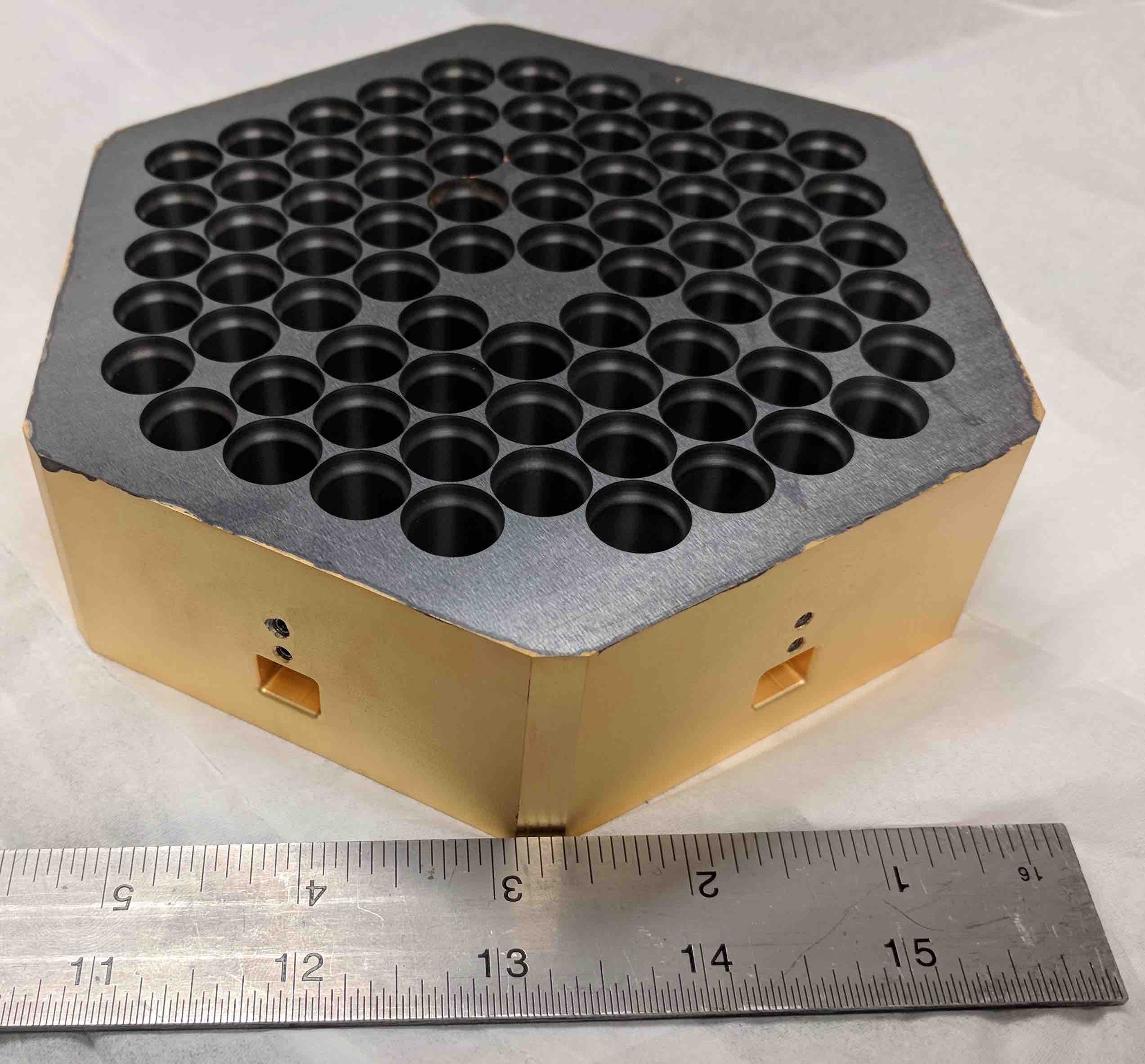}
\caption{The AdvACT 27/39~GHz feedhorn array is shown above. It is a direct-machined CE7F array. The AdvACT feedhorn array demonstrated 20~$\mathrm{\mu}$m tolerances and a feedhorn sidewall of 250~$\mathrm{\mu}$m.}
\label{fig:LF_AdvACT}
\end{figure}

CE7F is a Si-Al alloy produced by Sandvik Osprey\footnote{https://www.materials.sandvik/en/} that is planned for use on CLASS~\cite{Ali18} and the AdvACT 27/39~GHz array~\cite{Simon_2018}. The main advantage of CE7F over Al 6061 is that it has a $\Delta L/L=9\times 10^{-4}$~\cite{Ali18}, which is significantly closer to that of the Si detector array, resulting in less differential thermal contraction. However, CE7F is brittle like a ceramic, and the machining is currently proprietary. Because of its brittleness, CE7F requires helicoils for screw holes unlike other direct-machined metals. For the 27/39~GHz array, a bull nose cutter was used to machine the feedhorn profiles, and the inside of the feedhorns were not Au coated due to concerns about the uniformity of the Au coating across the feedhorn profile. The AdvACT feedhorn array shown in Fig.~\ref{fig:LF_AdvACT} demonstrated 20~$\mathrm{\mu}$m tolerances and a feedhorn sidewall of 250~$\mathrm{\mu}$m. An additional test piece demonstrated the capability to reach 150~$\mathrm{\mu}$m sidewalls. An initial test piece showed material pitting on the inside of the feedhorn profiles due to tool wear and tool overreach, but this was easily corrected by using longer tools and increasing the frequency of tool changes. The current CE7F cost at low frequencies is about 30\% the cost of a Si feedhorn array using photolithographic patterning. However, a 90/150~GHz array currently has equivalent pricing to stacked Si feedhorn array, and it is currently difficult to machine the required feedhorn sizes for the 220/270~GHz array in CE7F. The cost and feasibility at high frequency ruled out CE7F for SO, but Sandvik Osprey may develop the capabilities to produce feedhorn arrays at higher frequencies, increased production speed, and at reduced cost in the future.

\subsection{Molybdenum}
Mo has a $\Delta L/L=10.6\times 10^{-4}$, so it has less differential thermal contraction with the Si detector array than Al and similar differential thermal contraction as CE7F~\cite{White_78}. It also has a Vickers hardness ranging from 1400 to 2740~MPa compared to Al~6061, which has a value of 1049~MPa. While this increased hardness means that smaller walls between feedhorns could be possible, it also makes Mo more difficult to machine and wears down the custom tooling more rapidly than Al~6061. Additionally, the material itself is expensive compared to Al, and the custom tooling for Mo requires diamond coating, making it more expensive as well. The current cost per array is 25-67\% the cost of a Si feedhorn array using photolithographic patterning. The additional cost for Mo outweighed its benefits for SO, but it could be useful in future projects if less differential thermal contraction and smaller feedhorn walls were required.


\section{Silicon Feedhorn Arrays} \label{sec:silicon}
Si platelet feedhorn arrays are assembled by stacking individually etched Si wafers up to create the feedhorn profile~\cite{Nibarger_2012}. While Si arrays are more fragile and cannot have tapped mounting features directly machined into them, they have several advantages over direct-machined arrays. Because they are fabricated from the same material as the detector array, there are no differential thermal effects between the detector and feed arrays, which enables improved alignment between the pieces and less mechanical risk. The methods used to etch the Si wafers also have tighter tolerances than direct-machining methods, which is particularly important at high frequencies. The Si feedhorn array must be Au coated to make the Si feeds electrically conductive and to help with heat sinking. Unlike many metals used in direct-machining, Au coating Si arrays does not require any ferromagnetic materials. While we limit our discussion here to spline-profiled feedhorns, we note that because Si feedhorn arrays are built up from individual wafers, they can be used for more complex geometries.

To produce a Si feedhorn array, the wafers must first be etched with the hole patterns necessary to stack up the horn profile. The wafers are then coated in a metallic seed layer to enable even Au coating in the final step (AdvACT uses 200~nm of Ti and 1~$\mathrm{\mu}$m of Cu). Next the wafers are aligned and glued on the edges. NIST has demonstrated maximum misalignments between wafers of $<5\,\mathrm{\mu}$m with a maximum total misalignment of $<10\,\mathrm{\mu}$m across the entire feedhorn array. Finally, the assembled feedhorn array is electroplated in 3~$\mathrm{\mu}$m of Cu and 3~$\mathrm{\mu}$m of Au. The most time-consuming and labor-intensive step in the fabrication process is etching the wafers. In this section, we describe two methods for etching Si feedhorn array wafers: the current technique of etching using photolitographic patterning and laser machining.

\subsection{Etching with Photolithographic Patterning}
\begin{figure}[h!]
\centering
\includegraphics[width=0.6\textwidth]{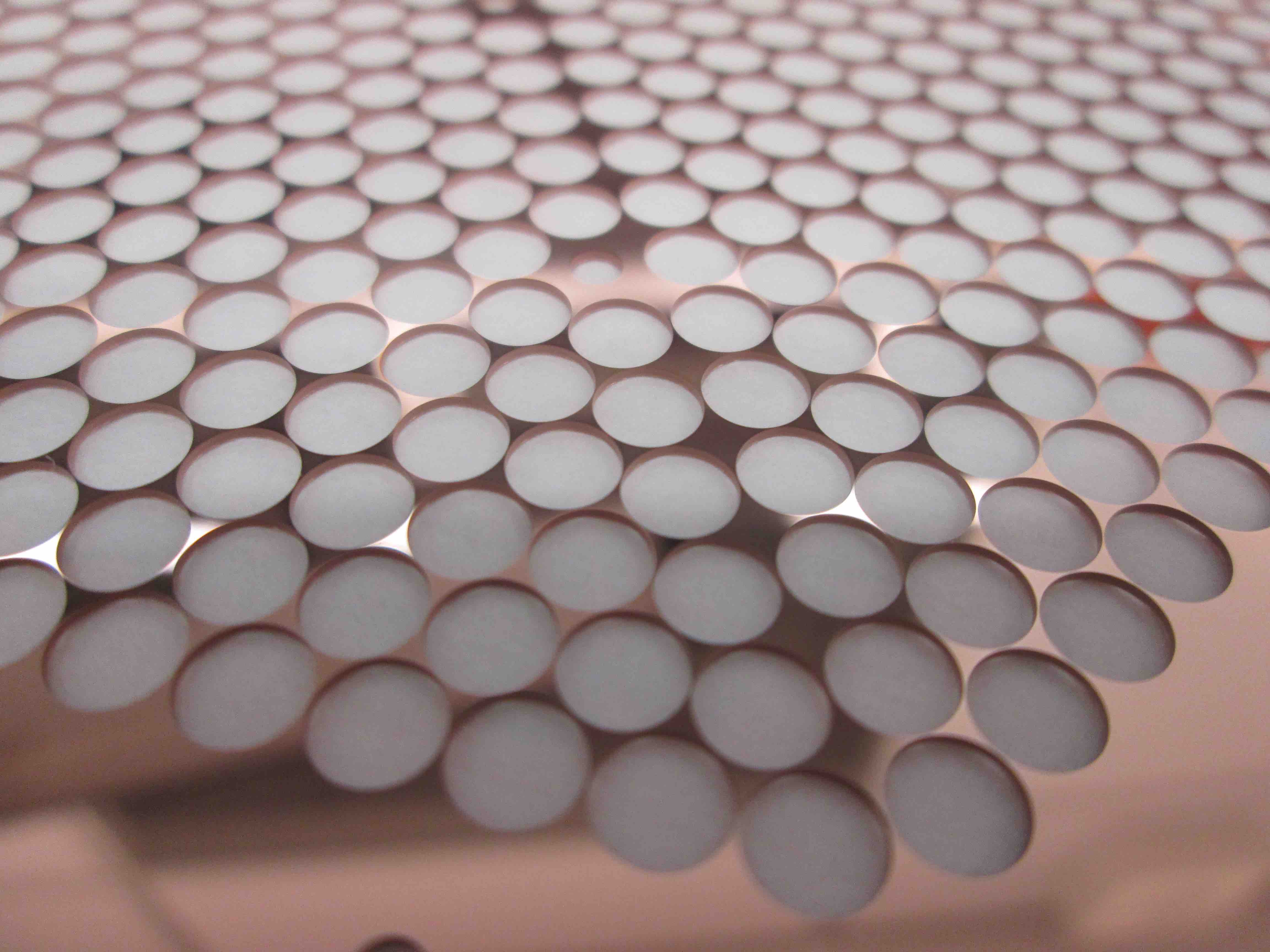}
\caption{The top wafer of the AdvACT 150/230~GHz feedhorn stack fabricated by NIST is shown above after it was coated with a seed layer of 200~nm Ti and 1~$\mathrm{\mu}$m of Cu. The feedhorn walls are 100~$\mathrm{\mu}$m on this wafer.}
\label{fig:single_wafer}
\end{figure}

\begin{figure}[h!]
\centering
\includegraphics[width=0.65\textwidth]{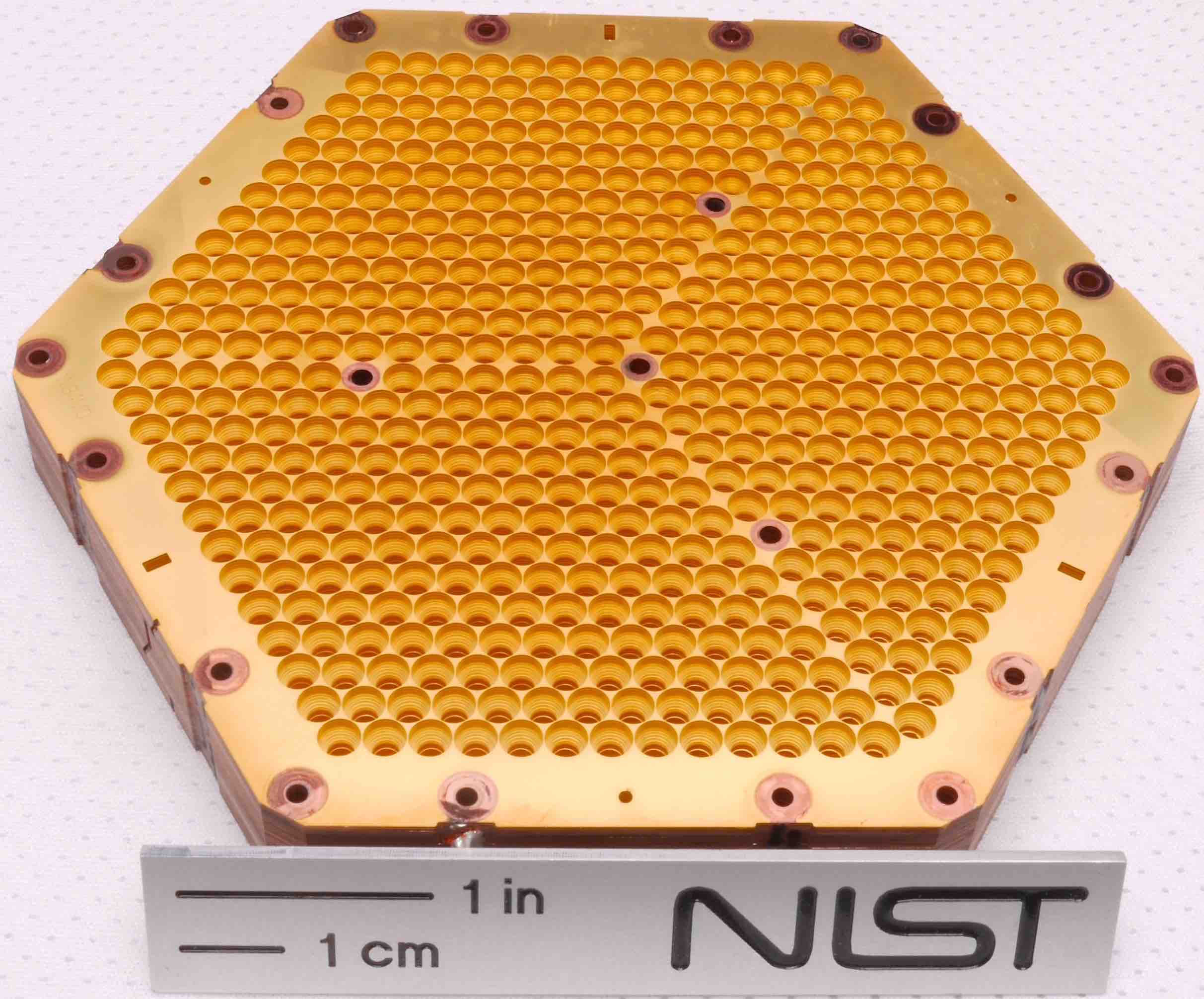}
\caption{The fully assembled AdvACT 150/230~GHz feedhorn array is shown above. To assemble the array, individual Si wafers are stacked up, aligned with dowel pins, and glued together. The wafers in this array are photolithographically-patterned and etched with a DRIE. Photo courtesy of Dan Schmidt.}
\label{fig:AdvACT_HF}
\end{figure}

NIST fabricates Si feedhorn arrays with photolitographic patterning on 150~mm Si wafers. Using photolithography and deep reactive ion etching (DRIE), each Si wafer has between one and three layers of the feedhorn profile etched into it. Photolitographic patterning can require fabricating a mask for each layer, and once the wafer is patterned, the holes in the wafer are etched with a DRIE. Figure~\ref{fig:single_wafer} shows a single layer of the AdvACT 150/230~GHz feedhorn array after its seed coating, and Fig.~\ref{fig:AdvACT_HF} shows the fully assembled feedhorn array.

NIST has demonstrated that the radial uncertainty with this method is $<2\mathrm{\mu}$m. The DRIE can create sidewall tapers that can be as large as 2$^{\circ}$ for a 500~$\mathrm{\mu}$m deep etch. However, the taper can be minimized by etching an annulus versus the whole area to remove the material. The smallest step size currently achievable with this method on a 150~mm wafer is 125~$\mathrm{\mu}$m, which is achieved by double-etching a 250-$\mathrm{\mu}$m-thick wafer. While wafers can have more than one layer with double and triple etching, this process requires additional cleaning and processing steps in the fabrication, increasing the required time and cost.

Photolithographic patterning has unparalleled tolerances, but the current fabrication process of even the simplest feedhorn array consisting of single-etched wafers is time-consuming, resulting in high costs. With the current process, the rate of production with two people in the fabrication lab is $\sim$5 single-etched wafers per day. To use stacked Si arrays on next-generation projects, the time and cost of etching the wafers must be drastically reduced.

\subsection{Laser Machining}
In laser machining, a laser is used to cut the features into each Si wafer. Laser machining a feedhorn array could allow for $\sim$300~mm feedhorn arrays, produce clean cuts, require no masks, and only take days to etch all the wafers for a full array, which will drastically decrease the cost of manufacturing. We estimate that this method could produce a wafer every hour with the work of only one person and have an equivalent cost to a direct-machined Al feedhorn array. We use a focusing lens with a long focal length to achieve a Rayleigh length of $\sim$350~$\mathrm{\mu}$m and to enable the ablation of material on micron scales. While DRIE-etching often leaves tapered sidewalls, the long Rayleigh length of the laser beam ensures that the laser-machined sidewalls are near-vertical, enabling a closer match to the feedhorn design. Laser machining also has the ability to etch multiple features into a single wafer without the additional processing steps required by DRIE etching. Multiple etches allow for higher resolution in the overall feedhorn design, which is particularly important at higher frequencies. The spot size and Rayleigh length can be tuned with different lenses to meet the requirements of a given etch. 

\begin{figure}[h!]
\centering
\includegraphics[width=0.3\textwidth]{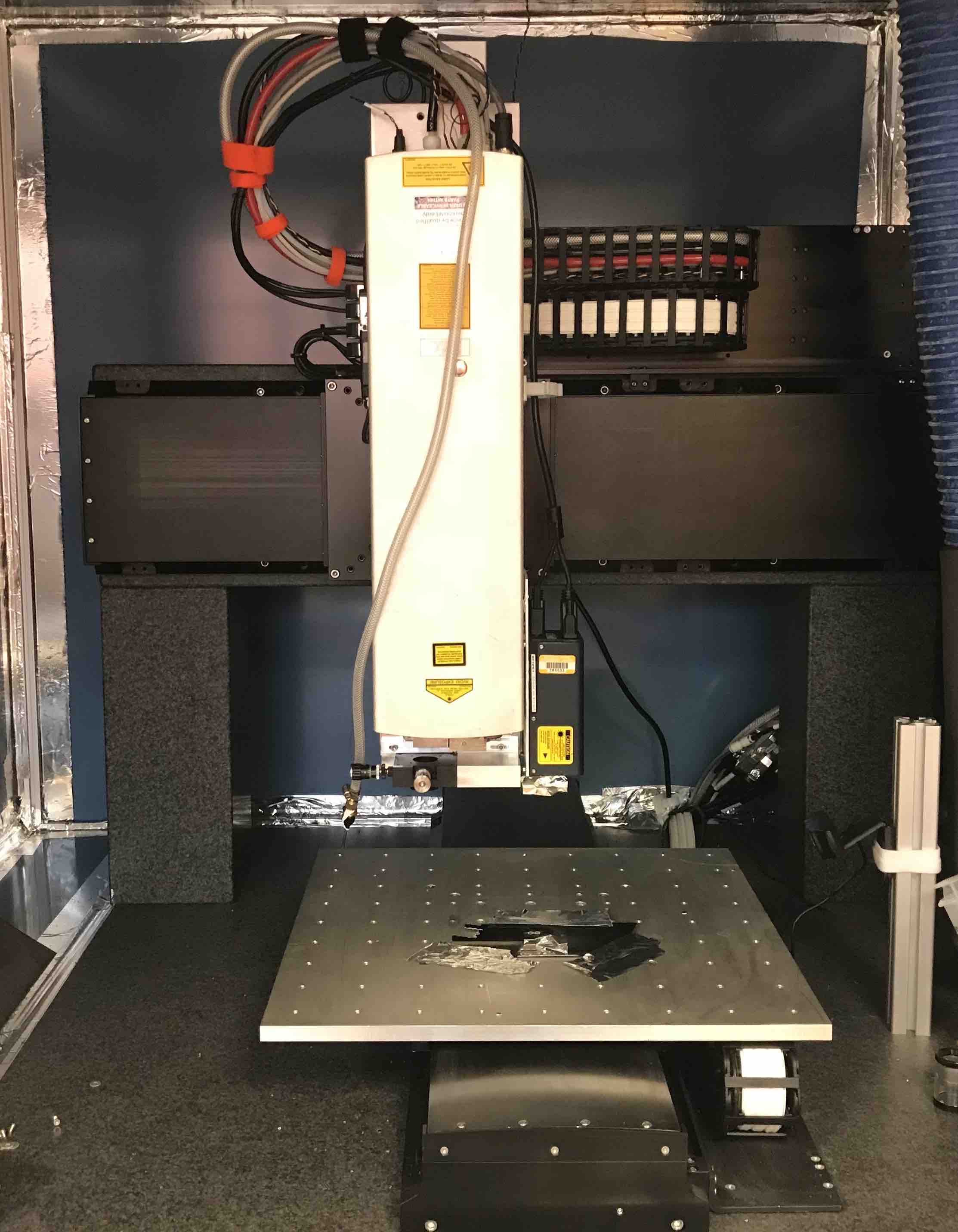}
\caption{The laser machining setup at the University of Michigan is shown above. The laser has a power of 20~W, and the stages have a speed of 60~mm/s with micron accuracy, allowing for large-format, rapid wafer production.}
\label{fig:laser_machining_setup}
\end{figure}

Our laser machining setup shown in Fig.~\ref{fig:laser_machining_setup} is a Talon 355-20 by Spectra Physics mounted on a custom three axis system produced by Aerotech. The laser is a green diode-pumped solid state Q-switched laser that produces 20~W of power when Q-switched at a repetition rate of 100~kHz. This power can be held over many repetitions within 0.3\% RMS, producing a high-powered, consistent laser beam ideal for micro-machining. The system stages have a 600~mm range of travel in both the x and y directions, so materials with a large area can be machined. We note that this could enable larger array sizes than 300~mm, but we are currently limited by the largest available Si wafer size of 300~mm. To allow for machining features at a range of heights, the laser is mounted on the z stage, which has a usable travel in the laser's focal range of $\sim$30~mm. The x and y stages move at a maximum rate of 60~mm/s and can locate and hold their positions with micron accuracy. The high movement speed and accuracy of the stages allows for the rapid production of near-micron accuracy features nearing the tolerances of photolithographic-patterning.

\begin{figure}[h!]
\centering
\includegraphics[width=0.95\textwidth]{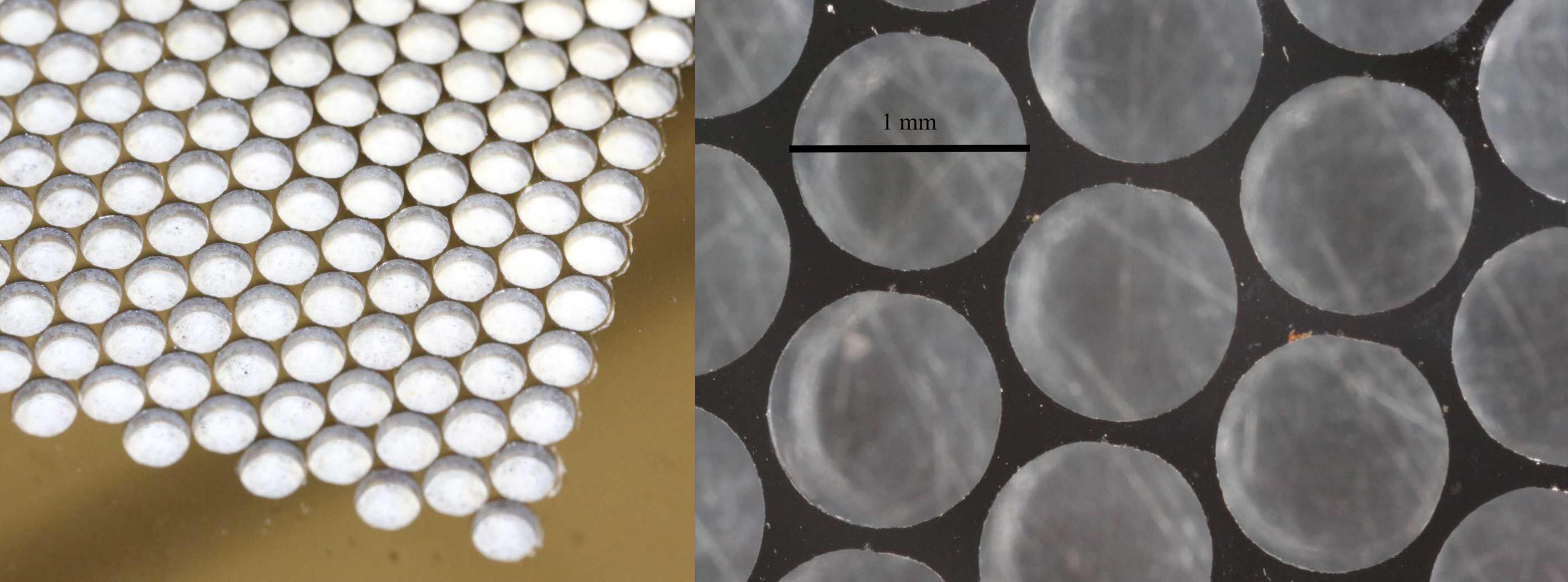}
\caption{The images above show a full wafer with a 1~mm hole size and 100~$\mathrm{\mu}$m sidewall. We note that the left image shows the full array before the removal of the Technikote. The sidewall is vertical, which presents an advantage over DRIE-etched holes, which can have a taper as large as 2$^{\circ}$. Laser machining also offers faster production speed and reduced cost for stacked Si detector arrays.}
\label{fig:laser_machined_wafer}
\end{figure}

To minimize and remove melt from the laser machining process, we flow high pressure nitrogen over the surface of the machined area. Argon was also considered, but we found no difference between argon and nitrogen. We also apply a thin layer of Technikote, a water soluble surface protectant, to protect the surface from redeposition. After the machining process is complete, the Technikote is rinsed off with warm water, resulting in a smooth surface with no redeposit present. Figure~\ref{fig:laser_machined_wafer} shows a laser-machined Si wafer with holes 1~mm in diameter and vertical sidewalls 100~$\mathrm{\mu}$m thick. This sample has little to no melt and there are no visible stress fractures from the heat of the laser. We are currently fabricating a single test feedhorn that we will compare to DRIE-etched feedhorns with beam maps.

We are upgrading our system with a galvanometer. The galvonometer will allow us to scan the laser spot over the surface at hundreds of millimeters per second. This will enable us to not only ablate material faster but will also give us greater control over how much material is removed, enabling better control of multi-layer etches. We estimate that this could enable etching a wafer every hour. The development schedule and advantages of laser machining will enable it to meet the production and science requirements of next-generation CMB experiments, making it a promising new technology.


\section{Summary and Outlooks}
SO plans to deploy multichroic feedhorn-coupled detector arrays at both 90/150~GHz and 220/270~GHz. Scaling up feedhorn production for SO and CMB-S4 requires drastic reductions in production time and cost. Direct-machining feedhorn arrays into metal enables low-cost, rapid production. While Si-Al alloy and Mo offer less differential thermal contraction with the Si detector array compared to Al~6061, they are currently limited by their cost and machinability. Al~6061 is low-cost and easy to machine, making it an ideal candidate for SO. However, direct machining has larger tolerances than photolithographically-patterned stacked Si feedhorn arrays, which could impact performance, particularly for the 220/270~GHz arrays. Profile measurements with a non-contact metrology probe and beam maps are underway to verify that Al feedhorns meet their tolerance requirements for SO.

Stacked Si feedhorn arrays offer high tolerances and no differential thermal contraction with the detector array. They have traditionally been patterned using photolithography and then DRIE-etched. This process can be long and costly. However, laser machining could significantly cut the production time and cost of stacked Si feedhorn arrays, making it a promising technology. We are developing both Al feedhorns and laser-machined stacked-Si feedhorns for SO, and both methods are promising for the cost and production needs of future CMB experiments.



\acknowledgments 
 
This work was supported in part by a grant from the Simons Foundation (Award \#457687, B.K.). We would like to thank Andrew Coleman of Sandvik Osprey and Tom Essinger-Hileman for many helpful discussions on the use of CE7F alloy.

\bibliography{report} 
\bibliographystyle{spiebib} 

\end{document}